\documentclass[prl,amsmath,amssymb,twocolumn,superscriptaddress,aps,floatfix,longbibliography]{revtex4-2}
\usepackage{graphicx}
\usepackage{dcolumn}
\usepackage{bm}
\usepackage{xcolor}
\usepackage{float}
\usepackage{standalone}
\usepackage{comment}
\usepackage[hidelinks]{hyperref}

 \graphicspath{ {./images/} }
 \usepackage{physics}

\begin{document}

\preprint{APS/123-QED}

\title{Band topology and dynamic multiferroicity induced from dynamical Dzyaloshinskii-Moriya interactions in centrosymmetric lattices}

\author{Bowen Ma}
\email{bowenphy@hku.hk}
\affiliation{HK Institute of Quantum Science \& Technology and Department of Physics, The University of Hong Kong, Pokfulam Road, Hong Kong, China}
\affiliation{Hong Kong Branch for Quantum Science Center of Guangdong-Hong Kong-Macau Great Bay Area, 3 Binlang Road, Shenzhen, China}

\author{Z. D. Wang}
\email{zwang@hku.hk}
\affiliation{HK Institute of Quantum Science \& Technology and Department of Physics, The University of Hong Kong, Pokfulam Road, Hong Kong, China}
\affiliation{Hong Kong Branch for Quantum Science Center of Guangdong-Hong Kong-Macau Great Bay Area, 3 Binlang Road, Shenzhen, China}

\date{\today}

\begin{abstract}
We develop a theory of a dynamical Dzyaloshinskii–Moriya interaction (dDMI) in centrosymmetric crystals by generally considering the vibration of both cations and anions. It gives rise to an antisymmetric spin-lattice coupling, inducing magnon-phonon hybridized topological excitations. Moreover, we find that this dDMI naturally exhibits a magnetoelectric feature, leading to the presence of dynamic multiferroicity with finite toroidal moment distribution in the momentum space. By comparing toroidal moments with band skyrmion structure, we reveal the intrinsic connection between band topology and dynamic multiferroicity through the dDMI. 
\end{abstract}

\maketitle

\emph{\textcolor{blue}{Introduction}}---The Dzyaloshinskii–Moriya interaction (DMI) is an antisymmetric exchange interaction between two localized spins. It was first proposed by Dzyaloshinskii~\cite{dzyaloshinsky1958thermodynamic} to phenomenologically explain the origin of the weak ferromagnetism in the antiferromagnet $\alpha$-Fe$_2$O$_3$. Later, Moriya~\cite{moriya1960new,moriya1960anisotropic} derived the spin Hamiltonian microscopically by using a second-order perturbation analysis of spin-orbit coupling (SOC) in Anderson's superexchange model. In recent decades, DMI has been intensively studied for understanding various physical phenomena including spin-induced multiferroicity~\cite{katsura2005spin,mostovoy2006ferroelectricity,katsura2007dynamical,sergienko2006role,kimura2003magnetic,ishiwata2008low},  non-reciprocal spin-wave propagation~\cite{zakeri2010asymmetric,udvardi2009chiral,moon2013spin,zivieri2018theory}, magnon thermal Hall effects~\cite{onose2010observation,laurell2018magnon,hwang2020topological}, spin torque and transport~\cite{manchon2014magnon,kovalev2016spin,ma2020longitudinal,lin2022evidence,yu2023voltage,ye2025spin}, and noncollinear spin textures such as chiral domain walls~\cite{heide2008dzyaloshinskii,thiaville2012dynamics,ryu2013chiral,sampaio2013nucleation} and magnetic skyrmions~\cite{mühlbauer2009skyrmion,yu2010real,jiang2015blowing,fert2017magnetic,li2021writing}. 
    
In addition to SOC, crystal symmetry has significant effects on inducing DMI. The orientation of the DM vector is restricted by the famous Moriya's rule~\cite{moriya1960anisotropic}, especially the local inversion symmetry within the bond $ij$ needs to be broken for the presence of a non-zero DM vector $\mathbf{D}_{ij}$. In centrosymmetric systems, by tuning lattice deformation, strain engineering has been demonstrated as a controllable method to manipulate the DMI~\cite{kitchaev2018phenomenology,davydenko2019dzyaloshinskii,zhang2021tunable} and to achieve different spin textures such as magnetic skyrmions~\cite{zhang2021strain,shen2022strain} and bimerons~\cite{cai2025stabilization}. However, from a dynamical point of view, even without a static distortion, lattice vibration can ``instantaneously'' break the inversional symmetry. This scenario is quite beyond the conventional Moriya's rule, and it is natural to ask if a DMI can then be induced from lattice dynamics even in a lattice with bond inversion symmetry.

With these motivations, in the present Letter, we generally propose a dynamical DMI (dDMI) from lattice vibrations in centrosymmetric lattices that couples spin degrees of freedom with lattice dynamics. Similar to the topological magnons induced from conventional static DMI~\cite{owerre2016topological,kim2016realization,mcclarty2022topological}, the excitation with the dDMI also develops non-trivial band topology. More remarkably, we notice the expression of the dDMI coincides with the spin-induced magnetoelectric effect in the Katsura-Nagaosa-Balatsky (KNB) model~\cite{katsura2005spin,katsura2007dynamical}, leading to dynamical entanglement of magnetization and electrical polarization. Similar dynamical effects have been studied in the literature as dynamic multiferroicity~\cite{juraschek2017dynamical,dunnett2019dynamic,juraschek2019dynamical,geilhufe2021dynamically,gao2023dynamical,basini2024terahertz,wang2024ultrafast,paiva2025dynamically} in non-magnetic systems. Since the KNB mechanism is responsible for type-II multiferroicity, we call this dynamical magnetoelectric feature of our dDMI model \textit{dynamic type-II} multiferroicity. By comparing the magnetoelectric toroidal moment~\cite{dubovik1990toroid,fiebig2005revival,spaldin2008toroidal,ederer2007towards} with the band skyrmion structure~\cite{qi2008topological,qi2011topological} in the momentum space, we find that the band topology and the dynamic type-II multiferroicity are naturally connected by the dDMI. Our work not only presents an unconventional effect of DMI in centrosymmetric lattices but also provides a magnetoelectric perspective to magnon-phonon couplings.
\begin{figure}
\includegraphics[width=0.42\textwidth]{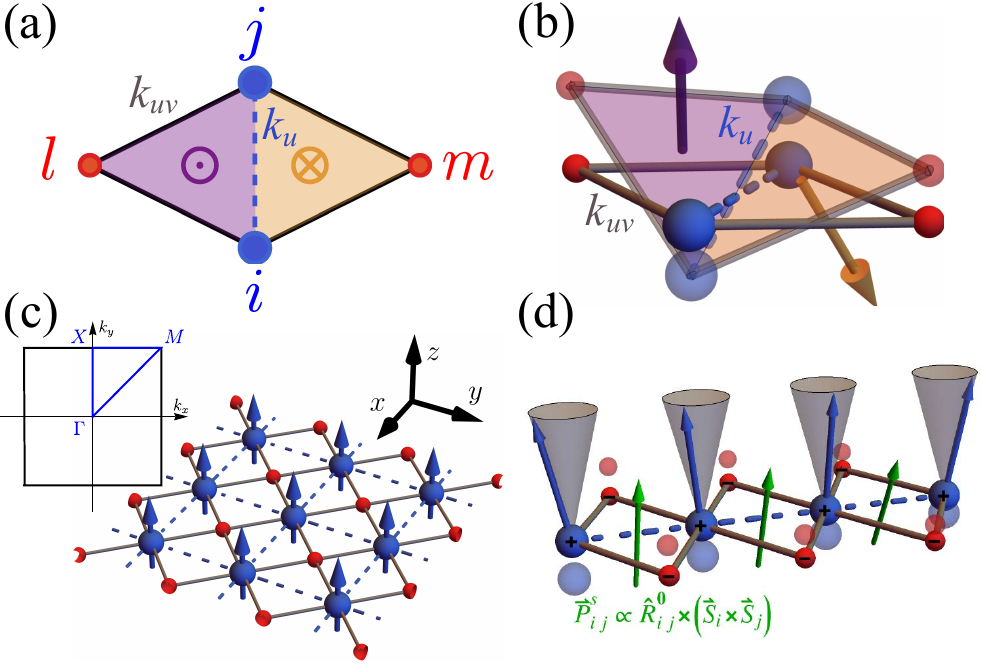}
    \caption{(a) Static four-atom diamond cell with $ij$ (blue) atoms magnetic but $lm$ (red) atom non-magnetic. (b) The schematic for dynamical DMI from lattice dynamics. $k_{u} (k_{uv})$ is the vibrational force constant between magnetic ions and nearest-neighbor (non-)magnetic ions. Purple and orange arrows indicate the DM vectors from purple and orange triangles, respectively. (c) A ferromagnet in a diatomic square lattice, where blue (red) spheres stand for (non-)magnetic ions. Blue arrows indicate magnetic moments along $z$-direction. The inset shows the Brillouin zone for the square lattice with a high-symmetry path $\Gamma XM\Gamma$. (d) The propagating spin-wave (blue arrows) produces varying polarizations (green arrows) $\mathbf{P}_{ij}^s$ that induce dynamical distortion of the lattice with electrical polarization $\mathbf{P}_{iljm}\propto(u_i^z+u_j^z-v_l^z-v_m^z)\hat{\mathbf{z}}$.} 
    \label{fig:Setup}
\end{figure}

\emph{\textcolor{blue}{Dynamical DMI}}---We begin from a four-atom diamond cell as shown in Fig.~\ref{fig:Setup}(a), where magnetic ions reside at site $\mathbf{R}_i$ and $\mathbf{R}_j$ with spin $\mathbf{S}_i$ and $\mathbf{S}_j$ respectively, while non-magnetic ligands are located at $\mathbf{R}_l$ and $\mathbf{R}_m$. Then, an out-of-plane DMI between $\mathbf{S}_i$ and $\mathbf{S}_j$ is allowed in principle, for example, in the (purple) triangular $\triangle_{ilj}$, as
\begin{align}
    H_{\triangle_{ilj}}\!=\!\mathbf{D}_{ilj}\!\cdot\!(\mathbf{S}_i\!\times\!\mathbf{S}_j),\ \text{with }\mathbf{D}_{ilj}\!=\!D_{ij}\frac{\mathbf{R}_{li}\!\times\!\mathbf{R}_{lj}}{|\mathbf{R}_{li}\!\times\!\mathbf{R}_{lj}|},
\end{align}
where $\mathbf{R}_{li(j)}=\mathbf{R}_{i(j)}-\mathbf{R}_l$ is the distance vector of the bond $i(j)l$, and $D_{ij}$ the coupling strength for the bond $ij$. The other (orange) triangle $\triangle_{imj}$ also contributes a DMI $\mathbf{D}_{imj}$ so that the total contribution is
\begin{align}
    H_D^0=(\mathbf{D}_{ilj}+\mathbf{D}_{imj})\cdot(\mathbf{S}_i\times\mathbf{S}_j).\label{HD}
\end{align}

Since this diamond structure is inversion symmetric at the bond-$ij$ center, according to Moriya's rule~\cite{moriya1960anisotropic,moriya1960new}, $\mathbf{D}_{ilj}$ and $\mathbf{D}_{imj}$ has the same coupling strength $D_{ij}$ but opposite direction, and thus the total DMI is exactly zero. Now if we consider the lattice vibration at finite temperature as a dynamical distortion, the perfect cancellation of the DMI is expected to be violated [See Fig.~\ref{fig:Setup}(b)]. To illustrate this idea, we introduce a small displacement $\mathbf{u}_{i(j)}$ of atoms $i(j)$ and $\mathbf{v}_{l(m)}$ of atoms $l(m)$ away from their equilibrium position $\mathbf{R}^0_{i(j)}$ and $\mathbf{R}^0_{l(m)}$ respectively. After expanding the interaction Eq.~\eqref{HD} to the first order in $\mathbf{u}$ and $\mathbf{v}$~\cite{sm}, we find a spin-lattice interaction as
\begin{align}
    H_{D}^{ij}\!=\!\frac{2D_0\!\tan\!\frac{\theta}{2}}{a}\left[\hat{\mathbf{R}}^0_{ij}\!\!\times\!\!\hat{\mathbf{z}}(\!v^z_l\!+\!v^z_m\!-\!u^z_i\!-\!u^z_j)\right]\!\!\cdot\!(\mathbf{S}_i\!\!\times\!\!\mathbf{S}_j),\label{dDMI}
\end{align}
where $D_0$ is the DMI from the triangle configuration in the equilibrium, $a=|\mathbf{R}^0_{ij}|$ is the lattice constant, $\theta$ is the angle between bond $li$ and $lj$ (or equivalently between bond $mi$ and $mj$) in the static limit, and $\hat{\mathbf{z}}$ is the direction perpendicular to the diamond plane. Due to its antisymmetric form of the coupling, this spin-lattice interaction can be regarded as a \textit{dynamical} DMI that couples the collective modes of spin-waves and lattice vibrations. In previous studies on DMI-induced magnon-phonon coupling~\cite{zhang2019thermal,ma2022antiferromagnetic}, a static non-zero DMI is needed with broken inversional symmetry from a heterostructure, and the strength of DMI needs to be small for not developing a spiral spin structure. However, in our case, the DMI is a dynamical one, which is originally hidden by the static symmetry, and thus will not undermine the classical magnetic order of the ground state.

\emph{\textcolor{blue}{Square lattice example}}---To study the physical consequences of this dDMI, we now apply the spin-lattice interaction Eq.~\eqref{dDMI} for the diamond cluster to a two-dimensional (2D) lattice system. In real materials, ligands may not be perfectly located in the magnetic lattice plane and $\hat{\mathbf{z}}$ for each diamond cluster can orient differently. For example in NiPS$_3$~\cite{wildes2015magnetic,kim2018charge} and CoTiO$_3$~\cite{yuan2020dirac,elliot2021order,choe2025long}, although the magnetic moment aligns in the magnetic lattice plane, and each diamond cluster is tilted from the plane, it still has a non-zero component perpendicular to each diamond plane. Therefore, for simplicity but without loss of generality, we consider a diatomic example in a 2D square lattice as shown in Fig.~\ref{fig:Setup}(c), where the ligands are sitting at each square center and the magnetic configuration is an out-of-plane ferromagnetic order. The minimal spin Hamiltonian can be written as
\begin{align}
    H_s=-J\sum_{\langle ij\rangle}\mathbf{S}_i\cdot\mathbf{S}_j-\sum_i\left[\eta(\mathbf{S}_i^z)^2+\mathcal{B}\mathbf{S}_i^z\right],\label{Hspin}
\end{align}
where $J>0$ is the nearest-neighbour ferromagnetic exchange coupling, $\eta$ is the single-ion anisotropy and $\mathcal{B}$ is the Zeeman splitting from external magnetic fields.

For this simple case, the effective dynamical DM vectors are in the plane and are all perpendicular to the collinear ferromagnetic order. Within the linear spin-wave regime, where spin operators are represented by creation and annihilation operators $a^\dagger$ and $a$ of Holstein-Primakoff magnons~\cite{holstein1940field} as $\mathbf{S}^\pm_i\approx\sqrt{2S}a_i (a_i^\dagger)$ and $\mathbf{S}^z_i=S-a^\dagger_ia_i^{}$, this DMI will not affect the spin dynamics. Nevertheless, the DM-type spin-lattice interaction can hybridize magnons with the collective modes of the lattice vibration if they are energetically close to each other~\cite{jensen1971magneto,vigren1972static}. These vibrational modes, i.e., phonons, can be described by an elastic Hamiltonian as
\begin{align}
    H_p&=\sum_i\frac{P_i^2}{2M}+\sum_l\frac{p_l^2}{2m}\nonumber\\
    &+\frac{1}{2}k_u\sum_{\langle\langle ij\rangle\rangle}(u^z_i-u^z_j)^2+\frac{1}{2}k_{uv}\sum_{\langle il\rangle}(u^z_i-v^z_l)^2,
\end{align}
where $P_i\ (p_l)$ is the momentum of (non-)magnetic atom at site $i\ (l)$ with mass $M\ (m)$, $k_{uv} (k_{u})$ is the first (second) nearest-neighbor force constant. Here we ignore the elastic coupling between non-magnetic ligands as they are usually not chemically bonded.

The full Hamiltonian is then $H=H_m+H_p+H_D$, and it can be Fourier transformed into the momentum space as a generalized Bogoliubov-de-Gennes (BdG) Hamiltonian $H=\frac{1}{2}\sum_\mathbf{k}\mathbf{X}_\mathbf{k}^\dagger H_\mathbf{k}^{}\mathbf{X}_\mathbf{k}^{}$ with $\mathbf{X}_\mathbf{k}^{}=(a_\mathbf{k}^{ }, a_\mathbf{-k}^\dagger, u^z_\mathbf{k}, v^z_\mathbf{k}, P_\mathbf{-k}^{ }, p_\mathbf{-k}^{ })^T$ and 
\begin{align}
    H_\mathbf{k}=\begin{pmatrix}
        \varepsilon^m_\mathbf{k} I_2 & H_{mp}(\mathbf{k}) & 0 & 0\\
        H^\dagger_{mp}(\mathbf{k}) & \Phi(\mathbf{k}) & 0 & 0\\
        0 & 0 & \frac{1}{M} & 0\\
        0 & 0 & 0 & \frac{1}{m}
    \end{pmatrix},
\end{align}
where $\varepsilon^m_\mathbf{k}=2JS[2-\cos (k_xa)-\cos (k_ya)]+B_\text{eff}$ is the pure magnon dispersion from Eq.~\eqref{Hspin} with the effective field $B_\text{eff}=\mathcal{B}+(2S-1)\eta$, $I_n$ is the $n\times n$ identity matrix, $\Phi(\mathbf{k})$ is the dynamical matrix of the vibration, and $H_{mp}(\mathbf{k})$ is the magnon-phonon coupling from the dDMI. The explicit expression of $\Phi(\mathbf{k})$ and $H_{mp}(\mathbf{k})$ is given in the Supplemental Material~\cite{sm}. 

\begin{figure}
\includegraphics[width=0.48\textwidth]{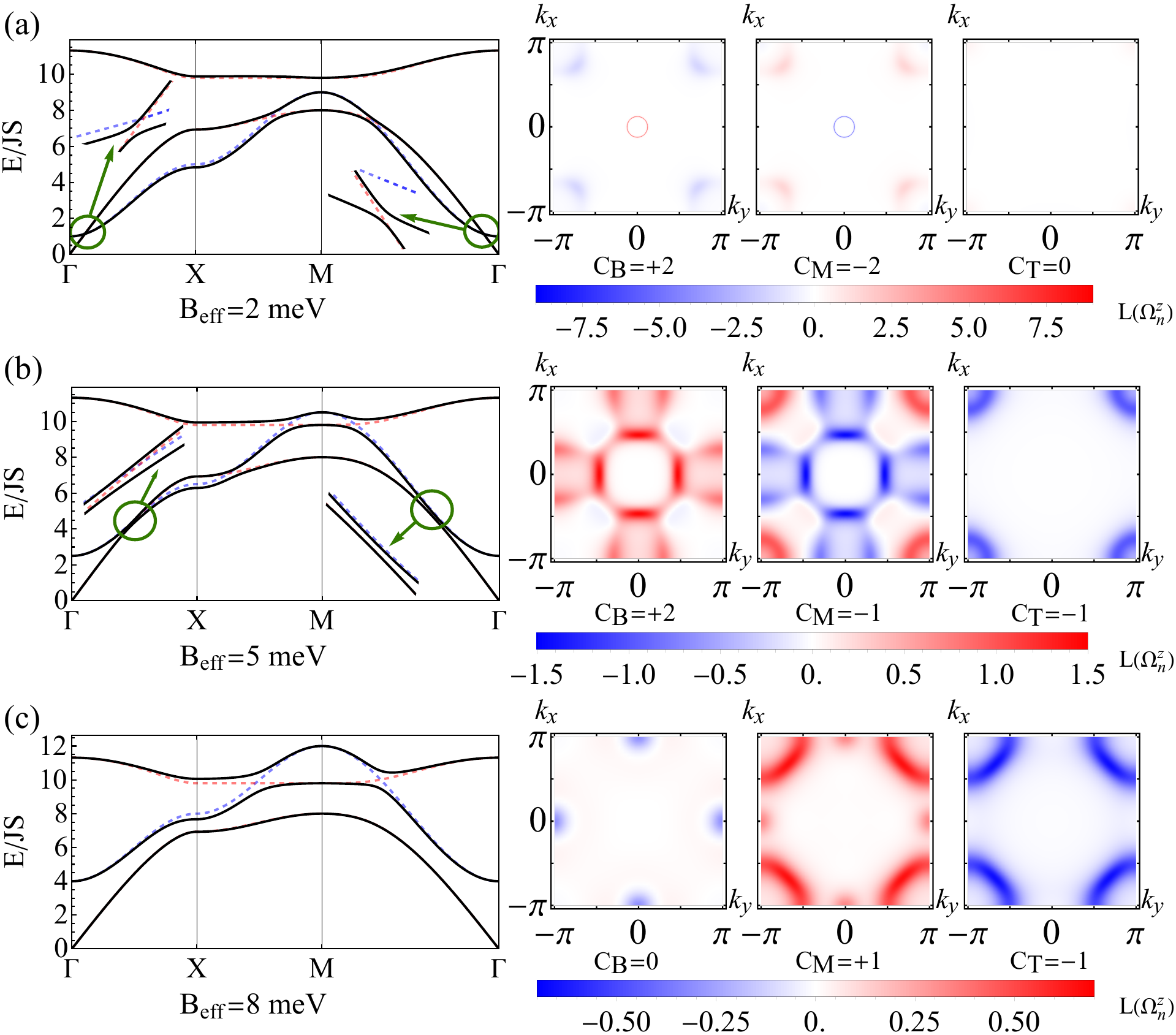}
    \caption{Band dispersion (left) and Berry curvature distribution (right) in log scale $L(\Omega_n^z)\equiv\text{sgn}(\Omega_n^z)\ln(1+|\Omega_n^z|)$ for (a) $B_\text{eff}=2$ meV, (b) $B_\text{eff}=5$ meV, (c) $B_\text{eff}=8$ meV. Red (Blue) dashed lines in the dispersion are uncoupled phonon (magnon) bands. $C_B, C_M,\text{ and }C_T$ is the Chern number for the bottom, middle, and top band, respectively.} 
    \label{fig:Band_topo}
\end{figure}

For this basis $\mathbf{X}_\mathbf{k}$, the commutator $[\mathbf{X}_\mathbf{k},\mathbf{X}_\mathbf{k}^\dagger]\equiv g$ is non-zero, and thus the $n$-th band dispersion $E_{n\mathbf{k}}$ of the hybrid excitations $|\psi_{n\mathbf{k}}\rangle$ can be calculated by the eigen-equation $g H_\mathbf{k}|\psi_{n\mathbf{k}}\rangle=E_{n\mathbf{k}}|\psi_{n\mathbf{k}}\rangle$~\cite{ma2024chiral} (see appendix A for details).
We depict the band dispersion in the left column of Fig.~\ref{fig:Band_topo} with $J=2$ meV, $S=1$ for typical 3d transition-metal compounds and take $m/M=1/3$ (e.g. O compared to Co approximately), $\omega_{u}\equiv\sqrt{k_u/M}=4$ meV, $\omega_{uv}\equiv\sqrt{k_{uv}/\tilde{M}}=8$ meV with $\tilde{M}\equiv\frac{2mM}{m+M}$. As the conventional DMI can be as large as $80\%$ of the Heisenberg exchange $J$~\cite{zhang2023giant}, we here set $D_0=1$ meV. With an external magnetic field, the magnonic band can be energetically manipulated to couple with acoustic and/or optical phonons by tuning $B_\text{eff}$, and a gap opens up at those crossings between magnon and phonon bands. Because optical modes are out-of-phase movements of atoms, they are more asymmetric than acoustic modes, and thus they lead to a more prominent gap opening when hybridizing with magnons by the dDMI.

\emph{\textcolor{blue}{Band topology}}---From the symmetry aspect, the dynamical DMI breaks the inversion symmetry, allowing non-trivial band topology. To show the topological effects of the DM-type spin-phonon coupling on the gap opening, we study the Berry curvature and band Chern numbers. For the BdG Hamiltonian, the Berry curvature $\boldsymbol{\Omega}_{n\mathbf{k}}$~\cite{berry1984quantal,takahashi2016berry,ma2022antiferromagnetic} and the corresponding band Chern number $C_n$ for the $n$-th band is defined as
\begin{align}
  \boldsymbol{\Omega}_{n\mathbf{k}}=i\boldsymbol{\nabla}_{\mathbf{k}}\times
\langle \psi_{n\mathbf{k}}|g\boldsymbol{\nabla}_\mathbf{k}|\psi_{n\mathbf{k}}\rangle,\ C_n=\frac{1}{2\pi}\sum_{\mathbf{k}}\boldsymbol{\Omega}^z_{n\mathbf{k}}.
\end{align}

In the right column of Fig.~\ref{fig:Band_topo}, we show the Berry curvature distribution in the momentum space and the Chern number for each band with different $B_\text{eff}$. It can be seen that large Berry curvature appears at those anti-crossing gaps where magnons and phonons are drastically hybridized by $H_{mp}$. The Chern numbers can be changed by moving the magnon band with an external magnetic field, which provides experimental tunability.

\begin{figure}
\includegraphics[width=0.4\textwidth]{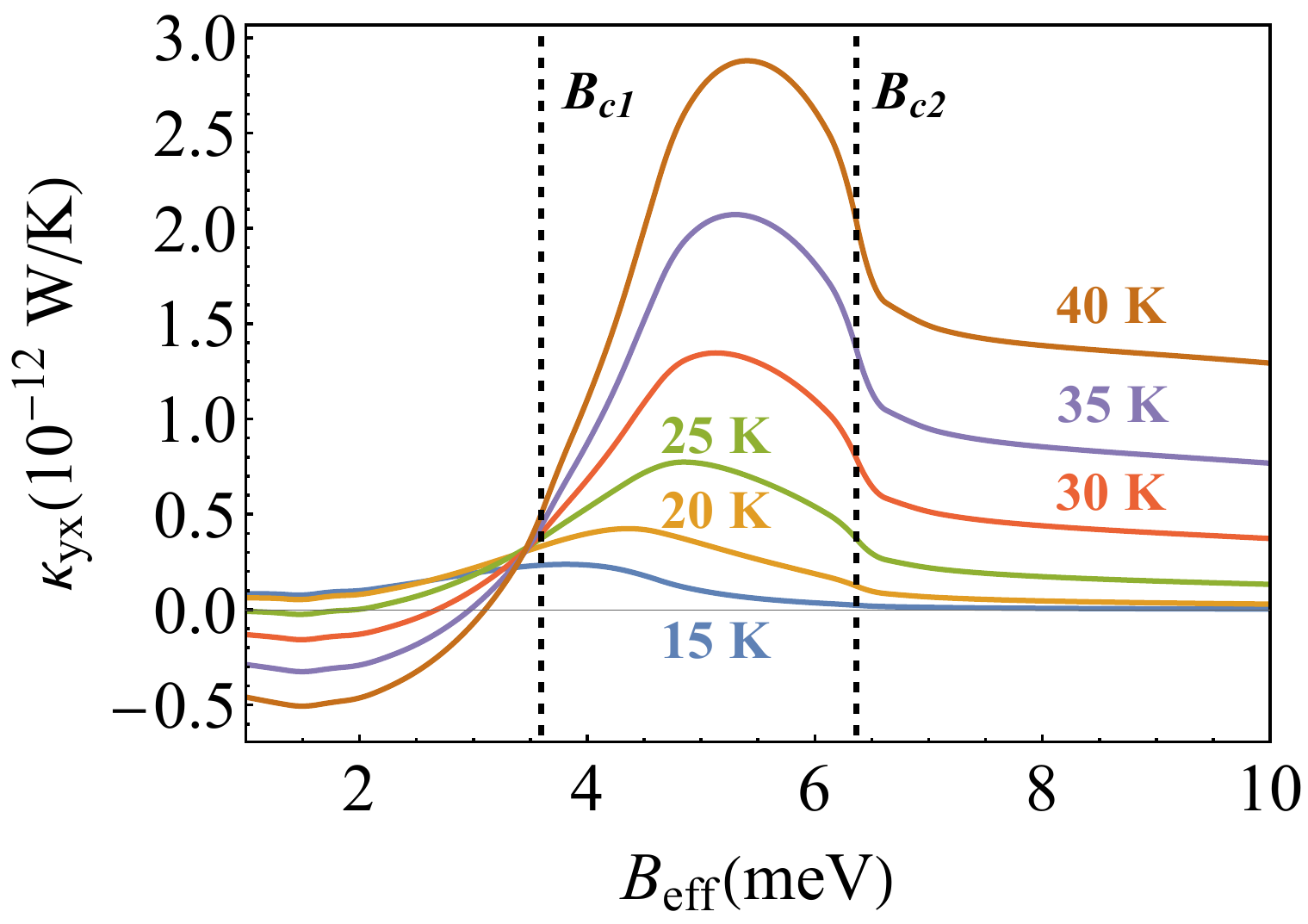}
    \caption{The dependence of thermal Hall effects on $B_\text{eff}$ for $T=15\text{ K},20\text{ K},25\text{ K},30\text{ K},35\text{ K, and }40\text{ K, respectively}$. The topological gap in Fig.~\ref{fig:Band_topo} closes at $B_{c1}\approx 3.60$ meV and $B_{c2}\approx 6.37$ meV.} 
    \label{fig:THC-B}
\end{figure}

As a powerful probe for detecting the topology of charge-neutral excitations, in our model Hamiltonian with the dDMI, we study the thermal Hall effects~\cite{strohm2005phenomenological,matsumoto2011theoretical,sugii2017thermal,grissonnanche2020chiral,zhang2024thermal}, where the thermal Hall conductivity is evaluated as~\cite{matsumoto2011rotational,matsumoto2011theoretical}
\begin{align}
\kappa_{yx}=\frac{k_B^2T}{\hbar V}\sum_{n,\mathbf{k}}\left[c_2(f(E_{n\mathbf{k}},T))-\frac{\pi^2}{3}\right]\Omega_{n\mathbf{k}}^z,
\end{align}
with $c_2(x)=(1+x)\ln^2(1+1/x)-\ln^2x-2\text{Li}_2(-x)$, Li$_2(x)$ is the polylogarithm function, $f$ the Bose-Einstein distribution function, $T$ is the average temperature, $V$ is the system volume. We estimate the Curie temperature is about $\frac{1}{3}S(S+1)zJ\sim60$ K with $z=4$ the coordination number~\cite{yosida1996theory}, and evaluate $\kappa_{yx}$ up to $40$ K in Fig.~\ref{fig:THC-B} with varying magnetic fields. The two vertical dashed lines represent the gap closing points separating the three topological phases in Fig.~\ref{fig:Band_topo},  and the trend of the curves clearly reflects these topological phase transitions under the effective magnetic field $B_\text{eff}$. We also estimate the longitudinal thermal conductivity and evaluate the thermal Hall angle in the appendix B.

\emph{\textcolor{blue}{Dynamic multiferroicity}}---Interestingly, although we derive Eq.~\eqref{dDMI} from a magnetoelastic consideration, the expression of the dDMI exhibits a magnetoelectric nature. This is because the electrical polarization in (ionic) crystals is associated with the vibration of charged ions. For example, in our diatomic model, cations and anions are supposed to have opposite effective charges $\pm Q^*$, and the vibration in the diamond $ilmj$ provides an electric polarization $\mathbf{P}_{iljm}\sim Q^*(u^z_i+u^z_j-v^z_l-v^z_m)\hat{\mathbf{z}}$ along $z$-direction. Therefore, the electric polarization $\mathbf{P}_{iljm}$ and the magnetization $\mathbf{M}_{ij}$ from spin $\mathbf{S}_i$ and $\mathbf{S}_j$ are coupled by the dDMI.

More remarkably, we can rewrite Eq.~\eqref{dDMI} as
\begin{align}
    H_{D}^{ij}\!=\!\frac{2D_0\!\tan\!\frac{\theta}{2}}{a}(u^z_i\!+\!u^z_j\!-\!v^z_l\!-\!v^z_m)\hat{\mathbf{z}}\!\cdot\!\!\left[\!\hat{\mathbf{R}}^0_{ij}\!\times\!\!(\!\mathbf{S}_i\!\!\times\!\!\mathbf{S}_j\!)\!\right].
\end{align}
According to the KNB mechanism~\cite{katsura2005spin}, $\hat{\mathbf{R}}^0_{ij}\!\times\!(\mathbf{S}_i\!\times\!\mathbf{S}_j)$ is proportional to an electrical polarization $\mathbf{P}^s_{ij}$ produced between $\mathbf{S}_i$ and $\mathbf{S}_j$. Then, as depicted in Fig.~\ref{fig:Setup}(d), the magnetoelectric entanglement of this dDMI can be understood in the spin-wave propagation picture~\cite{katsura2007dynamical,khomskii2009classifying}, where the precession of the spins induces a varying $\mathbf{P}^s_{ij}$ as well as an alternating electrical field coupling with $\mathbf{P}_{iljm}$. Consequently, despite the fact that the system is not ferroelectric as static polarization $\langle \mathbf{P}_{iljm}\rangle=0$, given that the fluctuation of the ferroelectric and the ferromagnetic degrees of freedom coexist and are entangled, this dDMI, as a magnetoelectric effect, will lead to the presence of dynamic type-II multiferroicity.

Since the band topology results from dDMI, the topological excitations should naturally possess multiferroic features. In the literature, to characterize multiferroicity,  the cross product of polarization $\mathbf{P}$ and magnetization $\mathbf{M}$, also known as toroidal moment $\mathbf{T}=\mathbf{P}\times\mathbf{M}$, has been introduced \cite{spaldin2008toroidal,arima2005resonant,sawada2005optical,kida2007optical}. In our case, a dynamical toroidal moment can be defined as,
\begin{align}
    \mathbf{T}&=\sum_{i}\left[Q^*(u^z_i-\frac{1}{4}\sum_{l\in n.n.}v^z_l)\hat{\mathbf{z}}\times\gamma\hbar\mathbf{S}_i\right]\equiv\sum_\mathbf{k}\mathbf{T}_\mathbf{k}
\end{align}
with $\gamma$ the gyromagnetic ratio, and $\mathbf{T}_\mathbf{k}$ the momentum-resolved toroidal moment. The explicit expression of $\mathbf{T}_\mathbf{k}$ is given in the Supplemental Material~\cite{sm}.

\begin{figure}
\includegraphics[width=0.48\textwidth]{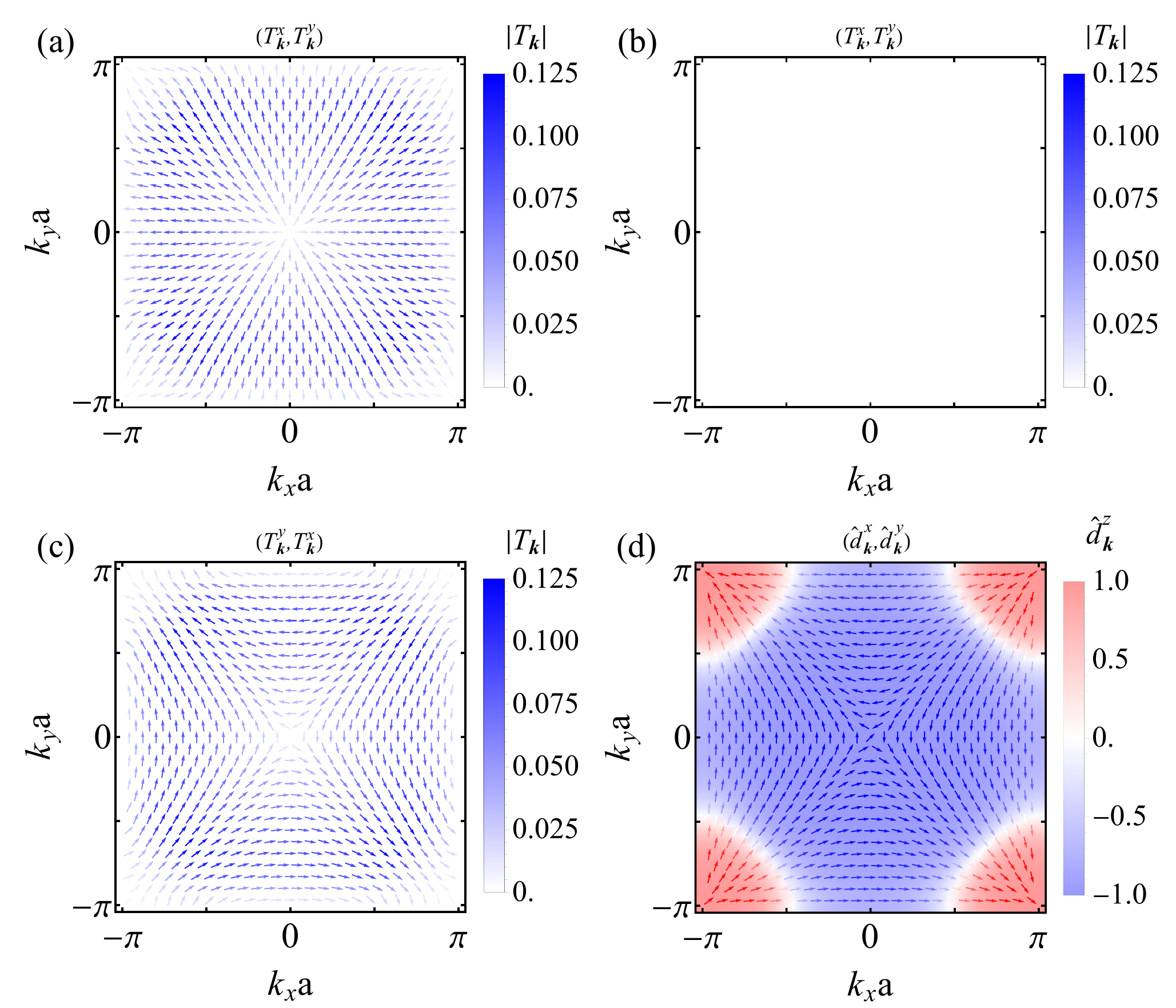}
    \caption{(a) Non-zero toroidal moment distribution (in the unit of $\hbar\gamma Q^*\sqrt{\frac{S}{2}}$) for the middle band for $B_\text{eff}=8$ meV when $D_0\neq 0$, characterizing dynamic type-II multiferroicity. (b) Toroidal moments become zero everywhere in the momentum space when $D_0=0$. (c) The rotated toroidal moment $(T^y_\mathbf{k},T^x_\mathbf{k})$, which is approximately proportional to $(\hat{d}^x_\mathbf{k},\hat{d}^y_\mathbf{k})$ as shown in Fig.~\ref{fig:TM}(d). (d) The skyrmion structure of the $\boldsymbol{d}$-vector for the effective magnon-phonon Hamiltonian.} 
    \label{fig:TM}
\end{figure}

As the effects of dDMI are more significant when magnons couple with optical phonons, we investigate the case with $B_\text{eff}=8$ meV as an example, where the acoustic phonon is energetically well-separated from the other two bands. In Fig.~\ref{fig:TM}(a) and (b), the finite and zero toroidal moment $\langle \psi_{2,\mathbf{k}}|\mathbf{T}_\mathbf{k}|\psi_{2,\mathbf{k}}\rangle$ for $D_0\neq 0$ and $D_0=0$ indicate that the dDMI is indeed necessary for inducing dynamic type-II multiferroicity. Since the acoustic phonon can be safely neglected from a perturbative point of view, to understand the relation between dDMI-induced band topology and the corresponding dynamic type-II multiferroicity, we focus on the effective optical phonon-magnon Hamiltonian written in the Pauli matrices $\boldsymbol{\sigma}=(\sigma_x,\sigma_y,\sigma_z)$ as
\begin{align}
    \tilde{H}_\mathbf{k}=\frac{1}{2}(\varepsilon^m_\mathbf{k}+\varepsilon^+_\mathbf{k})I_2+\boldsymbol{d}_\mathbf{k}\cdot\boldsymbol{\sigma}
\end{align}
with $\boldsymbol{d}_\mathbf{k}=(A_\mathbf{k}\sin{\frac{k_ya}{2}},A_\mathbf{k}\sin{\frac{k_xa}{2}},\frac{\varepsilon^m_\mathbf{k}-\varepsilon^o_\mathbf{k}}{2})$, where $\varepsilon_\mathbf{k}^{m(o)}$ is the dispersion of magnons (optical phonons) when $D_0=0$,  and $A_\mathbf{k}=\frac{4D_0\sqrt{S^3}}{a\sqrt{m\varepsilon^+_\mathbf{k}}}\left(\lambda\cos{\frac{k_ya}{2}}\sin{\theta_\mathbf{k}}+\cos{\frac{k_xa}{2}}\cos{\theta_\mathbf{k}}\right)$ with $\tan{2\theta_\mathbf{k}}=\frac{8\lambda\omega^2_{uv}\cos{\frac{k_xa}{2}}\cos{\frac{k_ya}{2}}}{4(1-\lambda^2)\omega^2_{uv}+(1+\lambda^2)(\cos{k_xa}+\cos{k_ya}-2)\omega^2_u}$ and $\lambda=\sqrt{m/M}$. The derivation details can be found in the Supplemental Material~\cite{sm}. Then, for the  two eigenstates $|\psi^\pm_\mathbf{k}\rangle$ of $\tilde{H}_\mathbf{k}$, it can be analytically derived that the corresponding toroidal moment $\langle\mathbf{T_\mathbf{k}}\rangle_\pm=\langle \psi^\pm_\mathbf{k}|\mathbf{T}_\mathbf{k}|\psi^\pm_\mathbf{k}\rangle$ is proportional to $\boldsymbol{d}$-vector as $\langle\mathbf{T_\mathbf{k}}\rangle_\pm=\mp c_\mathbf{k}\left(\hat{d}^y_{\mathbf{k}}\hat{\mathbf{x}}+\hat{d}^x_{\mathbf{k}}\hat{\mathbf{y}}\right)$, where $c_\mathbf{k}=\hbar\gamma Q^*\sqrt{\frac{S}{2}}\left(\frac{\sin\theta_\mathbf{k}}{\sqrt{2M\varepsilon^+_\mathbf{k}}}+\frac{\cos\theta_\mathbf{k}\cos{\frac{k_xa}{2}}\cos{\frac{k_ya}{2}}}{\sqrt{2m\varepsilon^+_\mathbf{k}}}\right)$, and $\hat{d}^\alpha_{\mathbf{k}}=d^\alpha_{\mathbf{k}}/|\boldsymbol{d}_\mathbf{k}|$ for $\alpha=x,y,z$. In Fig.~\ref{fig:TM}(c), we plot the distribution of $(T^y_\mathbf{k},T^x_\mathbf{k})$, which indeed matches with the pattern of the unit vector $\hat{\boldsymbol{d}}_\mathbf{k}$ in Fig.~\ref{fig:TM}(d).

On the other hand, the orientation of $\boldsymbol{d}$-vector over the $\mathbf{k}$-space reflects the topological property of the gap~\cite{qi2008topological,qi2011topological,bernevig2013topological}. In Fig.~\ref{fig:TM}(d), the unit vector $\hat{\boldsymbol{d}}_\mathbf{k}$ orientation exhibits a Neel-type skyrmion structure centered at $\mathbf{M}$ with a skyrmion number $Q_\text{skyrmion}=+1$. This is reminiscent of the winding number of $\boldsymbol{d}$-vector over the unit sphere in $\mathbf{k}$-space, and gives rise to the topological invariant Chern number $+Q_\text{skyrmion}(-Q_\text{skyrmion})$ for the lower (upper) band as the Berry curvature for $|\psi^{\pm}_\mathbf{k}\rangle$ is $\Omega^{\pm}_\mathbf{k}=\pm\frac{1}{2}\hat{\boldsymbol{d}}_\mathbf{k}\cdot\left(\partial_{k_x}\hat{\boldsymbol{d}}_\mathbf{k}\times\partial_{k_y}\hat{\boldsymbol{d}}_\mathbf{k}\right)$. Since $\hat{d}^x_{\mathbf{k}},\hat{d}^y_{\mathbf{k}}\propto D_0$, we find that the dDMI naturally connects hybridized band topology with dynamic multiferroicity characterized by the toroidal magnetoelectric moment.

\emph{\textcolor{blue}{Discussions}}---In this work, we used a ferromagnetic configuration as a proof of principle, but we emphasize that in the derivation of Eq.~\eqref{dDMI} there is no restriction on spin orientation, and thus our model can be generally applied to ferri-, antiferro-magnetic order or non-collinear spin textures as well. In particular, as antiferromagnets can host ultrafast spin dynamics in the terahertz frequency region~\cite{gomonay2018antiferromagnetic,baltz2018antiferromagnetic}, we expect a more notable effect from dDMI on the coupling between optical phonons and antiferromagnetic magnons~\cite{liu2021direct,cui2023chirality,luo2023evidence}.

Besides, the magnetoelectric nature of the dDMI suggests its possible application on multiferroic materials. One example is the 2D van der Waals multiferroic NiI$_2$~\cite{ju2021possible,song2022evidence}, which was recently suggested to be a type-II multiferroic material with a helimagnetic ground state. The conventional DMI is not allowed by the inversion symmetry~\cite{li2023realistic}, but it is interesting to consider the interplay between phonons and the helical order via the dDMI as well as the topology of their hybridization, which may provide electrical~\cite{shen2023electrical,neumann2024electrical,li2024electric} or optical~\cite{miyahara2012nonreciprocal,sato2016laser} approaches to exciting or detecting topological excitations. Given that the dDMI induces both finite Berry curvature and non-zero toroidal moments, similar to spin Nernst effects~\cite{zyuzin2016Magnon,cheng2016Spin,li2020intrinsic,ma2021Intrinsic}, transverse thermal transport of toroidal moments is expected to occur, which we leave for future study.

As the thermal Hall signal and gap opening induced from our theory are comparable to the results from other magnetoelastic models~\cite{go2019topological,zhang20203}, we expect our model to be complementary to them. In fact, our theory is primarily based on two intrinsic properties of materials, i.e., quantum phonon dynamics of non-magnetic ions and spin-orbital couplings, whose effects on magnon-phonon coupling have been observed in experiments~\cite{son2019unconventional} and evaluated in first-principle studies~\cite{fang2025efficient}. Thus, it is natural and necessary to include our mechanism, which was likely overlooked by the community in previous studies, to explain the experimental observation in centrosymmetric magnets, such as NiPS$_3$~\cite{meng2024thermodynamic} and FeCl$_2$~\cite{xu2023thermal}, especially for the high-temperature and/or high-field regime. 

In summary, we have proposed an unconventional dynamical DMI in centrosymmetric crystals by considering the lattice vibrations of both magnetic and non-magnetic atoms. It can hybridize magnon excitations with lattice dynamics and induce non-trivial band topology. More importantly, we remark on the intrinsic magnetoelectric feature of this dynamical DMI and introduce dynamic type-II multiferroicity via the well-known KNB model. By investigating the toroidal moment distribution and skyrmion structure in the momentum space, we further uncover that band topology and dynamic multiferroicity are essentially connected through dynamical DMI. We envision this study offers a new way for inducing and tuning topological excitations in various magnetic materials without inversion symmetry breaking, and suggests broad applications in spintronics, multiferroics, electromagnonics~\cite{xu2020floquet}, and magnetophononics~\cite{fechner2018magnetophononics}.

\begin{acknowledgments}
\emph{Acknowledgments}---We thank Ji Zou, Jeongheon Choe, Anyuan Gao, and Qian Niu for useful discussions, Gang v. Chen for inspiring references, and Gregory A. Fiete for constructive comments on the manuscript. This work is supported by the NSFC/RGC JRS grant with Grant No.~N$\_$HKU774/21, by the General Research Fund of Hong Kong with Grants No. 17310622 and No.~17303023.
\end{acknowledgments}
\emph{Notes added}---Upon preparing the manuscript, we became aware of a recent work~\cite{zhang2025spontaneous} where a dynamical spin-orbital coupling through phonon dynamics in centrosymmetric lattices is proposed in an electron hopping model. This microscopic model is compatible with our theory in principle and justifies the viability of our model. 


\bibliography{dDMI}

\appendix
\onecolumngrid
\begin{center}
  \textbf{\large End Matter}
\end{center}
\twocolumngrid

\setcounter{equation}{0}
\setcounter{figure}{0}
\setcounter{table}{0}
\makeatletter
\renewcommand{\theequation}{A\arabic{equation}}
\renewcommand{\thefigure}{A\arabic{figure}}
\renewcommand{\bibnumfmt}[1]{[A#1]}

\emph{Appendix A: Generalized BdG Hamiltonian}---In this appendix, we explain the diagonalization of the magnon-phonon coupled BdG Hamiltonian Eq.~(6) in the main-text.

We begin from a general bosonic BdG Hamiltonian for $N$ particles
\begin{align}
    H=\frac{1}{2}\sum_{\mathbf{k}}\mathbf{X}_{\mathbf{k}}^\dagger H_{\mathbf{k}}\mathbf{X}_{\mathbf{k}},\label{HX}
\end{align}
with a 2$N$ dimensional basis $\mathbf{X}_{\mathbf{k}}$ satisfying a commutator relation
\begin{align}
    [\mathbf{X}^{}_{\mathbf{k}}, \mathbf{X}_{\mathbf{k}}^\dagger]=g.\label{gX}
\end{align}
The basis $\mathbf{X}_\mathbf{k}$ can be transformed into a bosonic representation $\mathbf{Y}_\mathbf{k}=U_\mathbf{k}^{-1}\mathbf{X}_\mathbf{k}$, which diagonalize the Hamiltonian as
\begin{align}
    H=\frac{1}{2}\sum_{\mathbf{k}}\mathbf{Y}_{\mathbf{k}}^\dagger \mathbf{E}_{\mathbf{k}}\mathbf{Y}_{\mathbf{k}},\label{Y}
\end{align}
where $\mathbf{E}_\mathbf{k}=\text{diag}(E_{1\mathbf{k}},E_{2\mathbf{k}},...,E_{2N\mathbf{k}})$ is a diagonal matrix.
It should be noted that $\mathbf{Y}_\mathbf{k}$ needs to satisfy the bosonic commutator in the particle-hole space as 
\begin{align}
    [\mathbf{Y}_{\mathbf{k}}, \mathbf{Y}_{\mathbf{k}}^\dagger]=\begin{pmatrix}
    I_N & 0\\
    0 & -I_N
\end{pmatrix}\equiv\sigma_3.\label{sigma3}
\end{align}
From Eq.~\eqref{gX}, Eq.~\eqref{Y} and Eq.~\eqref{sigma3}, one can obtain
\begin{align}
    &g=U_\mathbf{k}[\mathbf{Y}_\mathbf{k}, \mathbf{Y}_\mathbf{k}^\dagger]U_\mathbf{k}^\dagger=U_\mathbf{k}\sigma_3U_\mathbf{k}^\dagger\Rightarrow U_\mathbf{k}^\dagger=\sigma_3U_\mathbf{k}^{-1}g,\\
    &U_\mathbf{k}^\dagger H_\mathbf{k}U_\mathbf{k}=\mathbf{E}_\mathbf{k}\Rightarrow g H_\mathbf{k}U_\mathbf{k}=U_\mathbf{k}\sigma_3\mathbf{E}_\mathbf{k},\label{XY}
\end{align}
where the $n$-th column of $U_\mathbf{k}$ corresponds to the linear representation under basis $\mathbf{X}_\mathbf{k}$ for eigenvector $\left|\psi_{n\mathbf{k}}\right>$ in the main text. Therefore, the eigen-equation for the BdG Hamiltonian is written as $g H_\mathbf{k}|\psi_{n\mathbf{k}}\rangle=E_{n\mathbf{k}}|\psi_{n\mathbf{k}}\rangle$ for $n=1,2,...,N$. In the main-text, $\mathbf{X}_\mathbf{k}^{}=(a_\mathbf{k}^{ }, a_\mathbf{-k}^\dagger, u^z_\mathbf{k}, v^z_\mathbf{k}, P_\mathbf{-k}^{ }, p_\mathbf{-k}^{ })^T$, so that the commutator can be obtained as
\begin{align}
    g=[\mathbf{X}^{}_\mathbf{k},\mathbf{X}_\mathbf{k}^\dagger]=\begin{pmatrix}
    1 & 0 & 0 & 0\\
    0 & -1&  0 & 0\\
    0 & 0 & 0&iI_2\\
    0& 0 & -iI_2 & 0
\end{pmatrix}.
\end{align}

\setcounter{equation}{0}
\setcounter{figure}{0}
\setcounter{table}{0}
\makeatletter
\renewcommand{\theequation}{B\arabic{equation}}
\renewcommand{\thefigure}{B\arabic{figure}}
\renewcommand{\bibnumfmt}[1]{[B#1]}
\begin{figure}
    \centering
    \includegraphics[width=0.44\textwidth]{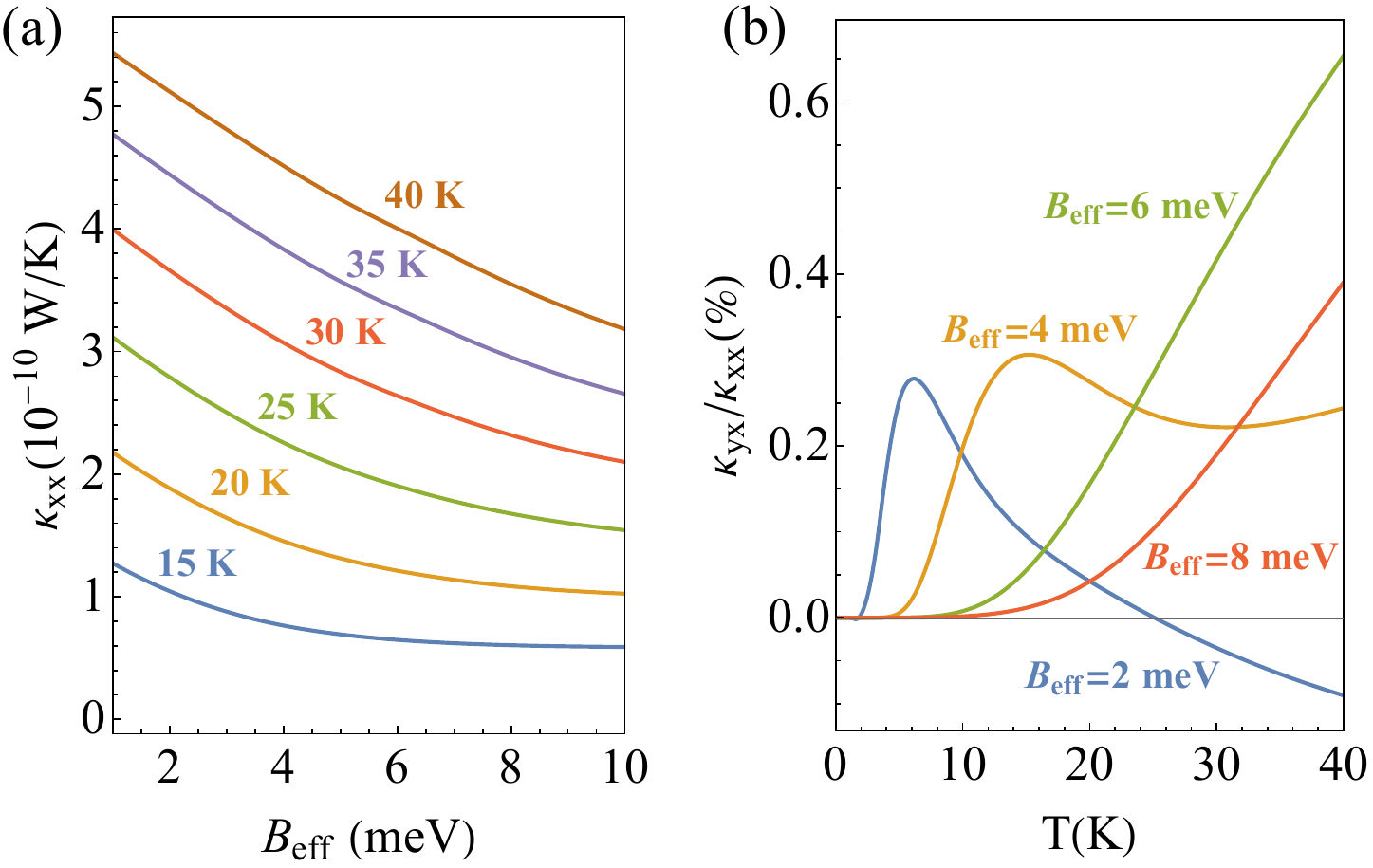}
    \caption{(a) The dependence of longitudinal thermal conductivity $\kappa_{xx}$ on $B_\text{eff}$
for $T = 15$ K, $20$ K, $25$ K, $30$ K, $35$ K, and $40$ K, respectively. (b) The dependence of the thermal Hall angle on temperature $T$
for $B_{\text{eff}} = 2$ meV, $4$ meV, $6$ meV, $8$ meV, respectively.}
    \label{fig:HA}
\end{figure}
\emph{Appendix B: Thermal Hall angle}---As the thermal Hall angle, i.e., the ratio between $\kappa_{yx}$ and longitudinal thermal conductivity $\kappa_{xx}$,  is an important quantity in the experiments to evaluate the thermal Hall effects, in this section, we give an estimate for $\kappa_{xx}$ and evaluate thermal Hall angle $\theta_\text{th}=\kappa_{yx}/\kappa_{xx}$. Theoretically, the longitudinal coefficient depends on the relaxation time $\tau$ and can be calculated as~\cite{ogasawara2025longitudinal},
\begin{align}
    \kappa_{xx}(T)=-\frac{1}{TV}\sum_{n,\mathbf{k}}\tau \mathbf{v}_{n\mathbf{k}}^2[E_{n\mathbf{k}}-\mu]^2\frac{\partial f(E_{n\mathbf{k}},T)}{\partial E_{n\mathbf{k}}},
\end{align}
where $E_{n\mathbf{k}}, \mathbf{v}_{n\mathbf{k}}\equiv\hbar^{-1}\nabla_\mathbf{k}E_{n\mathbf{k}}, \mu$ and $f(E_{n\mathbf{k}},T)$ are the energy dispersion, group velocity, chemical potential and Bose-Einstein distribution function respectively. Generally, the relaxation time $\tau$ depends on the scattering process and is a function of  $T, \mathbf{k}$ and $E_{n\mathbf{k}}$. Its calculation requires approximations and empirical formulae with several free parameters~\cite{walker1963phonon,bayrakci2013lifetimes}, but for the purpose of an estimation, we use $\tau$ as a constant. For phonon-phonon scattering and magnon-magnon scattering, the relaxation time is typically within the order of $10^{-1}-10^2$ ps, and thus we take $\tau=1$ ps. The results are shown in Fig.~\ref{fig:HA}. In general, the relaxation time at low (high) temperature should be longer (shorter), and thus we expect a higher (lower) thermal Hall angle in the experiments.

\clearpage
\onecolumngrid
\begin{center}
\textbf{\large Supplementary Materials for ``Band topology and dynamic multiferroicity induced from dynamical Dzyaloshinskii-Moriya interactions in centrosymmetric lattices''}
\end{center}
\begin{center}
    Bowen Ma$^{1,2}$, and Z. D. Wang$^{1,2}$
\end{center}
\begin{center}
    \textit{\small $^1$HK Institute of Quantum Science \& Technology and Department of Physics,
The University of Hong Kong, Pokfulam Road, Hong Kong, China\\
$^2$Hong Kong Branch for Quantum Science Center of Guangdong-Hong Kong-Macau Great Bay Area, 3 Binlang Road, Shenzhen, China}
\end{center}

\setcounter{equation}{0}
\setcounter{figure}{0}
\setcounter{table}{0}
\setcounter{page}{1}
\makeatletter
\renewcommand{\theequation}{S\arabic{equation}}
\renewcommand{\thefigure}{S\arabic{figure}}
\renewcommand{\bibnumfmt}[1]{[S#1]}
\renewcommand{\citenumfont}[1]{S#1}

\section{DM-like spin-lattice coupling}
In this section, we derive $H_D$ in Eq.~(3). We write the displacement for the atoms from their equilibrium position $\mathbf{R}^0_i$, $\mathbf{R}^0_j$, $\mathbf{R}^0_l$, $\mathbf{R}^0_m$ as $\mathbf{u}_i\equiv \mathbf{R}_i-\mathbf{R}^0_i$, $\mathbf{u}_j\equiv \mathbf{R}_j-\mathbf{R}^0_j$, $\mathbf{v}_l\equiv \mathbf{R}_l-\mathbf{R}^0_l$, and $\mathbf{v}_m\equiv \mathbf{R}_m-\mathbf{R}^0_m$. At this point, they are not limited to moving in $\hat{\mathbf{z}}$-direction but have all three translational degrees of freedom. For notation convenience, we define $\mathbf{u}_{ij}\equiv \mathbf{u}_j-\mathbf{u}_i$, $\mathbf{u}_{il}\equiv \mathbf{v}_l-\mathbf{u}_i$, $\mathbf{u}_{jl}\equiv \mathbf{v}_l-\mathbf{u}_j$, $\mathbf{u}_{im}\equiv \mathbf{v}_m-\mathbf{u}_i$, and $\mathbf{u}_{jm}\equiv \mathbf{v}_m-\mathbf{u}_j$.

We first look into the direction of $\mathbf{D}_{ilj}$,
\begin{align}
    \frac{\mathbf{R}_{li}\times\mathbf{R}_{lj}}{|\mathbf{R}_{li}\times\mathbf{R}_{lj}|}&=\frac{(\mathbf{R}^0_{li}+\mathbf{u}_{li})\times(\mathbf{R}^0_{lj}+\mathbf{u}_{lj})}{|(\mathbf{R}^0_{li}+\mathbf{u}_{li})\times(\mathbf{R}^0_{lj}+\mathbf{u}_{lj})|}=\frac{(\mathbf{R}^0_{li}+\mathbf{u}_{li})\times(\mathbf{R}^0_{lj}+\mathbf{u}_{lj})}{\sqrt{[(\mathbf{R}^0_{li}+\mathbf{u}_{li})\times(\mathbf{R}^0_{lj}+\mathbf{u}_{lj})]\cdot[(\mathbf{R}^0_{li}+\mathbf{u}_{li})\times(\mathbf{R}^0_{lj}+\mathbf{u}_{lj})]}}\nonumber\\
    &=\frac{\mathbf{R}^0_{li}\times\mathbf{R}^0_{lj}+\mathbf{R}^0_{li}\times\mathbf{u}_{lj}+\mathbf{u}_{li}\times \mathbf{R}^0_{lj}+\dots}{|\mathbf{R}^0_{li}\times\mathbf{R}^0_{lj}|}\left[1+\frac{2(\mathbf{R}^0_{li}\times\mathbf{R}^0_{lj})\cdot(\mathbf{R}^0_{li}\times\mathbf{u}_{lj}+\mathbf{u}_{li}\times \mathbf{R}^0_{lj})+\dots}{|\mathbf{R}^0_{li}\times\mathbf{R}^0_{lj}|^2}\right]^{-\frac{1}{2}}\nonumber\\
    &\approx\frac{\mathbf{R}^0_{li}\times\mathbf{R}^0_{lj}}{|\mathbf{R}^0_{li}\times\mathbf{R}^0_{lj}|}\left[1-\frac{(\mathbf{R}^0_{li}\times\mathbf{R}^0_{lj})\cdot(\mathbf{R}^0_{li}\times\mathbf{u}_{lj}+\mathbf{u}_{li}\times \mathbf{R}^0_{lj})}{|\mathbf{R}^0_{li}\times\mathbf{R}^0_{lj}|^2}\right]+\frac{\mathbf{R}^0_{li}\times\mathbf{u}_{lj}+\mathbf{u}_{li}\times \mathbf{R}^0_{lj}}{|\mathbf{R}^0_{li}\times\mathbf{R}^0_{lj}|}.
\end{align}

For the magnitude part, in general, it could depend on the bond length $|\mathbf{R}_{li}|$, $|\mathbf{R}_{lj}|$, and $\mathbf{R}_{li}\cdot\mathbf{R}_{lj}$ as the angle between them, while even if we only consider its dependence on $|\mathbf{R}_{ij}|$, it is adequate for us to obtain the non-trivial results. Besides, for the small displacement, the assumption that $D_{ij}$ depends only on the bond length $|\mathbf{R}_{ij}|=R$ is qualitatively valid to capture the effects of lattice vibration, i.e.,
\begin{align}
    D_{ij}=D(|\mathbf{R}_{ij}|)=D\left(\sqrt{(\mathbf{R}^0_{ij}+\mathbf{u}_{ij})^2}\right)\approx D(|\mathbf{R}^0_{ij}|)+\frac{\partial D(R)}{\partial R}\frac{\mathbf{R}^0_{ij}\cdot \mathbf{u}_{ij}}{|\mathbf{R}^0_{ij}|}
\end{align}

Combining Eq.~(S1) and Eq.~(S2), we obtain $\mathbf{D}_{ilj}$ to the linear order of the small displacement $\mathbf{u}$ and $\mathbf{v}$ as
\begin{align}
    \mathbf{D}_{ilj}&\approx D(|\mathbf{R}^0_{ij}|)\left\{\frac{\mathbf{R}^0_{li}\times\mathbf{R}^0_{lj}}{|\mathbf{R}^0_{li}\times\mathbf{R}^0_{lj}|}\left[1-\frac{(\mathbf{R}^0_{li}\times\mathbf{R}^0_{lj})\cdot(\mathbf{R}^0_{li}\times\mathbf{u}_{lj}+\mathbf{u}_{li}\times \mathbf{R}^0_{lj})}{|\mathbf{R}^0_{li}\times\mathbf{R}^0_{lj}|^2}\right]+\frac{\mathbf{R}^0_{li}\times\mathbf{u}_{lj}+\mathbf{u}_{li}\times \mathbf{R}^0_{lj}}{|\mathbf{R}^0_{li}\times\mathbf{R}^0_{lj}|}\right\}\nonumber\\
    &+\frac{\partial D(R)}{\partial R}\frac{\mathbf{R}^0_{ij}\cdot \mathbf{u}_{ij}}{|\mathbf{R}^0_{ij}|}\frac{\mathbf{R}^0_{li}\times\mathbf{R}^0_{lj}}{|\mathbf{R}^0_{li}\times\mathbf{R}^0_{lj}|},
\end{align}
and $\mathbf{D}_{imj}$ can be simply obtained by changing the sub-index $l$ to $m$.

Noticing $\mathbf{R}^0_{mj}=-\mathbf{R}^0_{li}$, $\mathbf{R}^0_{mi}=-\mathbf{R}^0_{lj}$, $\mathbf{u}_{il}+\mathbf{u}_{jm}=\mathbf{u}_{jl}+\mathbf{u}_{im}$ and $|\mathbf{R}^0_{mj}|=|\mathbf{R}^0_{li}|=|\mathbf{R}^0_{mi}|=|\mathbf{R}^0_{lj}|\equiv R_0$, we have
\begin{align}
    \mathbf{D}_{ij}=\mathbf{D}_{ilj}+\mathbf{D}_{imj}=D(|\mathbf{R}^0_{ij}|)\left[\frac{\mathbf{R}_{ij}^0\times(\mathbf{u}_{il}+\mathbf{u}_{jm})}{|\mathbf{R}^0_{li}\times\mathbf{R}^0_{lj}|}+\frac{\mathbf{R}^0_{li}\times\mathbf{R}^0_{lj}}{|\mathbf{R}^0_{li}\times\mathbf{R}^0_{lj}|}\frac{(R_0^2-\mathbf{R}^0_{li}\cdot\mathbf{R}^0_{lj})(\mathbf{R}^0_{li}+\mathbf{R}^0_{lj})\cdot(\mathbf{u}_{il}+\mathbf{u}_{jm})}{|\mathbf{R}^0_{li}\times\mathbf{R}^0_{lj}|^2}\right].
\end{align}

Now we define $D(|\mathbf{R}^0_{ij}|)\equiv D_0$, $\frac{\mathbf{R}^0_{li}\times\mathbf{R}^0_{lj}}{|\mathbf{R}^0_{li}\times\mathbf{R}^0_{lj}|}\equiv\hat{\mathbf{z}}$, $|\mathbf{R}^0_{ij}|\equiv a$, $\frac{\mathbf{R}^0_{ij}}{|\mathbf{R}^0_{ij}|}\equiv \hat{\mathbf{R}}^0_{ij}$, $\frac{a}{2R_0}=\sin\frac{\theta}{2}$, and notice that $\mathbf{R}^0_{li}+\mathbf{R}^0_{lj}=\mathbf{R}^0_{lm}=\frac{\mathbf{R}^0_{ij}\times \hat{\mathbf{z}}}{\tan\frac{\theta}{2}}$, we then have
\begin{align}
    \mathbf{D}_{ij}=\frac{2D_0}{a}\tan\frac{\theta}{2}\left[\hat{\mathbf{R}}^0_{ij}\times(\mathbf{u}_{il}+\mathbf{u}_{jm})+\hat{\mathbf{z}}(\hat{\mathbf{R}}^0_{ij}\times\hat{\mathbf{z}})\cdot(\mathbf{u}_{il}+\mathbf{u}_{jm})\right].
\end{align}
Since $\hat{\mathbf{R}}^0_{ij}=\hat{\mathbf{z}}\times(\hat{\mathbf{R}}^0_{ij}\times\hat{\mathbf{z}})$, we finally get the form of the DMI as
\begin{align}
    \mathbf{D}_{ij}&=\frac{2D_0}{a}\tan\frac{\theta}{2}\left\{\left[\hat{\mathbf{z}}\times(\hat{\mathbf{R}}^0_{ij}\times\hat{\mathbf{z}})\right]\times(\mathbf{u}_{il}+\mathbf{u}_{jm})+\hat{\mathbf{z}}(\hat{\mathbf{R}}^0_{ij}\times\hat{\mathbf{z}})\cdot(\mathbf{u}_{il}+\mathbf{u}_{jm})\right\}\nonumber\\
    &=\frac{2D_0}{a}\tan\frac{\theta}{2}(\hat{\mathbf{R}}^0_{ij}\times\hat{\mathbf{z}})\left[(\mathbf{u}_{il}+\mathbf{u}_{jm})\cdot\hat{\mathbf{z}}\right]=\frac{2D_0\tan\frac{\theta}{2}}{a}\left[\hat{\mathbf{R}}^0_{ij}\times\hat{\mathbf{z}}(v^z_l+v^z_m-u^z_i-u^z_j)\right],
\end{align}
which only involves the out-of-plane displacement and does not depend on the elastic coefficient $\frac{\partial D(R)}{\partial R}$.
Therefore, for each diamond $\diamond iljm$, the DMI within the bond $ij$ between the two magnetic atoms $i$ and $j$ is 
\begin{align}
    H_{D}^{ij}&=\frac{2D_0\tan\frac{\theta}{2}}{a}\left[\hat{\mathbf{R}}^0_{ij}\times\hat{\mathbf{z}}(v^z_l+v^z_m-u^z_i-u^z_j)\right]\cdot(\mathbf{S}_i\times\mathbf{S}_j)\nonumber\\
    &=\frac{2D_0\tan\frac{\theta}{2}}{a}\left[\left(\hat{\mathbf{R}}^0_{ij}\cdot\mathbf{S}_i\right)\left(\hat{\mathbf{z}}\cdot\mathbf{S}_j\right)-\left(\hat{\mathbf{R}}^0_{ij}\cdot\mathbf{S}_j\right)\left(\hat{\mathbf{z}}\cdot\mathbf{S}_i\right)\right](v^z_l+v^z_m-u^z_i-u^z_j).
\end{align}
For ferromagnetic cases, to the lowest order, $\mathbf{S}_i\approx \sqrt{\frac{S}{2}}\left(a^{}_i+a^\dagger_i\right)\hat{\mathbf{x}}+i\sqrt{\frac{S}{2}}\left(a^\dagger_i-a^{}_i\right)\hat{\mathbf{y}}+S\hat{\mathbf{z}}$. Then, in the Fourier space, we have magnon-phonon coupling $H_{mp}\approx \sum_{\langle ij \rangle}H_{D}^{ij}$ as
\begin{align}
    H_{mp}&=\frac{2D_0\tan{\frac{\theta}{2}}\sqrt{2S^3}}{a}\sum_\mathbf{k}\sum_{\diamond{ilmj}}\left[\left(\hat{\mathbf{R}}^0_{ij}\cdot\hat{\mathbf{y}}\right)\left(a^\dagger_\mathbf{k}-a^{}_{-\mathbf{k}}\right)-i\left(\hat{\mathbf{R}}^0_{ij}\cdot\hat{\mathbf{x}}\right)\left(a^\dagger_\mathbf{k}+a^{}_{-\mathbf{k}}\right)\right]\nonumber\\
&\times\left(u^z_{\mathbf{k}}\sin{\mathbf{k}_{ij}}-2v^z_{\mathbf{k}}\sin{\frac{\mathbf{k}_{ij}}{2}}\cos{\frac{\mathbf{k}_{lm}}{2}}\right),
\end{align}
where $\mathbf{k}_{ij(lm)}\equiv\mathbf{k}\cdot\mathbf{R}^0_{ij(lm)}$, and the summation is over all diamonds in the unit cell.

In the simple square lattice example, $\theta=\pi/2$, and $\hat{\mathbf{R}}^0_{ij}=\{\hat{\mathbf{x}},\hat{\mathbf{y}}\}$. Then, we have
\begin{align}
    H_{mp}&=\frac{2D_0\sqrt{2S^3}}{a}\sum_\mathbf{k}\left(a^\dagger_\mathbf{k}-a^{}_{-\mathbf{k}}\right)\left[\sin{(k_y a)}u^z_{\mathbf{k}}-2\sin{\frac{k_ya}{2}}\cos{\frac{k_x a}{2}}v^z_{\mathbf{k}}\right]\nonumber\\
    &-i\frac{2D_0\sqrt{2S^3}}{a}\sum_\mathbf{k}\left(a^\dagger_\mathbf{k}+a^{}_{-\mathbf{k}}\right)\left[\sin{(k_x a)}u^z_{\mathbf{k}}-2\sin{\frac{k_xa}{2}}\cos{\frac{k_y a}{2}}v^z_{\mathbf{k}}\right],
\end{align}
and $H_{mp}(\mathbf{k})$ in Eq.~(6) can be explicitly written as
\begin{align}
    H_{mp}(\mathbf{k})=\frac{2D_0\sqrt{2S^3}}{a}\begin{pmatrix}
 \sin {k_ya}-i\sin{k_xa} & -2 \cos{\frac{k_xa}{2}} \sin{\frac{k_ya}{2}}+2i \sin{\frac{k_xa}{2}} \cos {\frac{k_ya}{2}}\\
 -\sin {k_ya}-i\sin {k_xa} & 2 \cos{\frac{k_xa}{2}} \sin{\frac{k_ya}{2}}+2i \sin {\frac{k_xa}{2}}\cos {\frac{k_ya}{2}}
    \end{pmatrix}.
\end{align}

\begin{figure}
    \centering
    \includegraphics[width=0.75\textwidth]{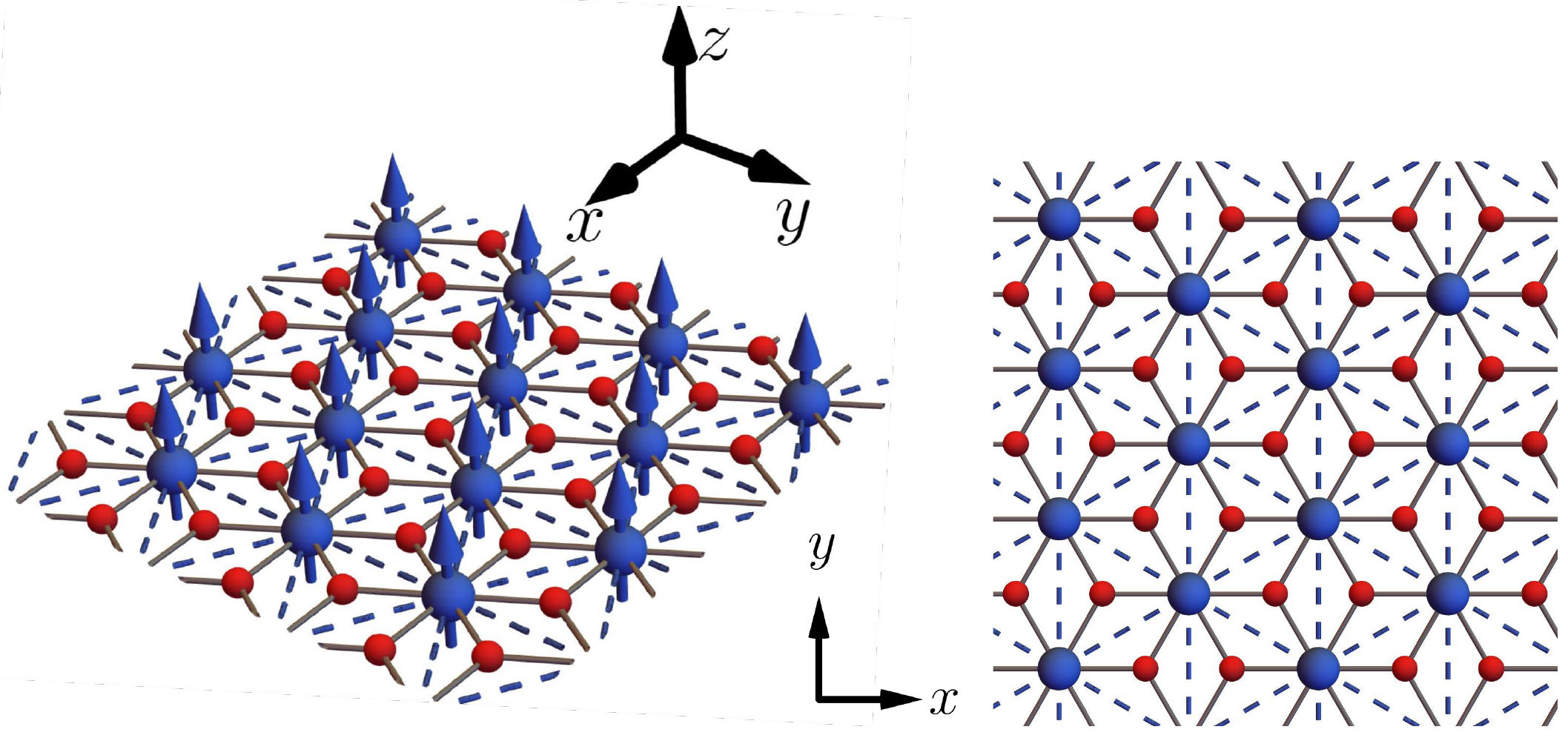}
    \caption{A ferromagnet in a triangular lattice, where blue (red) spheres stand for (non-)magnetic ions. Blue arrows indicate magnetic moments along $z$-direction.}
    \label{fig:TL}
\end{figure}

As some candidate materials such as FeCl$_2$ and NiI$_2$ belong to trigonal crystals, we consider the magnon-phonon coupling via dDMI in a triangular lattice as shown in Fig.~\ref{fig:TL}.
In this case, $\theta=2\pi/3$, and $\hat{\mathbf{R}}^0_{ij}=\{\frac{\sqrt3}{2}\hat{\mathbf{x}}-\frac{1}{2}\hat{\mathbf{y}},-\frac{\sqrt3}{2}\hat{\mathbf{x}}-\frac{1}{2}\hat{\mathbf{y}},\hat{\mathbf{y}}\}$. Then, we have
\begin{align}
    H_{mp}&=\frac{2\sqrt{3}D_0\sqrt{2S^3}}{a}\sum_\mathbf{k}\left(a^\dagger_\mathbf{k}-a^{}_{-\mathbf{k}}\right)\left[\left(\cos{\frac{\sqrt{3}k_x a}{2}}\sin{\frac{k_ya}{2}}+\sin{k_y a}\right)u^z_{\mathbf{k}}-3\cos{\frac{k_x a}{2\sqrt{3}}}\sin{\frac{k_ya}{2}}v^z_{\mathbf{k}}\right]\nonumber\\
    &-i\frac{6D_0\sqrt{2S^3}}{a}\sum_\mathbf{k}\left(a^\dagger_\mathbf{k}+a^{}_{-\mathbf{k}}\right)\left[\sin{\frac{\sqrt{3}k_xa}{2}}\cos{\frac{k_ya}{2}}u^z_{\mathbf{k}}-\left(\sin\frac{k_x a}{\sqrt{3}}+\sin\frac{k_x a}{2\sqrt{3}}\cos\frac{k_ya}{2}\right)v^z_{\mathbf{k}}\right].
\end{align}
Therefore, our theory can also be applied to centrosymmetric triangular magnets.

\section{Phonon-magnon effective model}
In this section, we derive the effective model for the optical phonon-magnon coupling studied in the main text. We first investigate the phononic part generally,
\begin{align}
    H_p=\sum_\mathbf{k}\frac{P_{\mathbf{k}}P_{-\mathbf{k}}}{2M}+\frac{p_{\mathbf{k}}p_{-\mathbf{k}}}{2m}+\frac{1}{2}(u^z_{-\mathbf{k}}, v^z_{-\mathbf{k}})\Phi_\mathbf{k}
    \begin{pmatrix}
        u^z_{\mathbf{k}}\\v^z_{\mathbf{k}}
    \end{pmatrix}.
\end{align}
In this simple case of the $2\times 2$ dynamical matrix $\Phi(\mathbf{k})$, this phonon Hamiltonian can be analytically diagonalized as $H_p=\sum_\mathbf{k}\varepsilon_\mathbf{k}^+b^\dagger_{\mathbf{k}}b^{}_{\mathbf{k}}+\varepsilon_\mathbf{k}^-\bar{b}^\dagger_{\mathbf{k}}\bar{b}^{}_{\mathbf{k}}$, where
\begin{align}
    &u^z_\mathbf{k}=\frac{\bar{b}^\dagger_{-\mathbf{k}}+\bar{b}^{}_\mathbf{k}}{\sqrt{2M\varepsilon^-_\mathbf{k}}}\cos\theta_\mathbf{k}-\frac{b^\dagger_{-\mathbf{k}}+b^{}_\mathbf{k}}{\sqrt{2M\varepsilon^+_\mathbf{k}}}\sin\theta_\mathbf{k},\quad P_{-\mathbf{k}}=i\sqrt{\frac{M\varepsilon^-_\mathbf{k}}{2}}\left(\bar{b}^\dagger_{-\mathbf{k}}-\bar{b}^{}_\mathbf{k}\right)\cos\theta_\mathbf{k}-i\sqrt{\frac{M\varepsilon^+_\mathbf{k}}{2}}\left(b^\dagger_{-\mathbf{k}}-b^{}_\mathbf{k}\right)\sin\theta_\mathbf{k},\\
    &v^z_\mathbf{k}=\frac{\bar{b}^\dagger_{-\mathbf{k}}+\bar{b}^{}_\mathbf{k}}{\sqrt{2m\varepsilon^-_\mathbf{k}}}\sin\theta_\mathbf{k}+\frac{b^\dagger_{-\mathbf{k}}+b^{}_\mathbf{k}}{\sqrt{2m\varepsilon^+_\mathbf{k}}}\cos\theta_\mathbf{k},\quad p_{-\mathbf{k}}=i\sqrt{\frac{m\varepsilon^-_\mathbf{k}}{2}}\left(\bar{b}^\dagger_{-\mathbf{k}}-\bar{b}^{}_\mathbf{k}\right)\sin\theta_\mathbf{k}+i\sqrt{\frac{m\varepsilon^+_\mathbf{k}}{2}}\left(b^\dagger_{-\mathbf{k}}-b^{}_\mathbf{k}\right)\cos\theta_\mathbf{k},
\end{align}
with $\theta_\mathbf{k}=\frac{1}{2}\arctan\left[\frac{2\Phi_{2\mathbf{k}}}{\Phi_{1\mathbf{k}}-\Phi_{3\mathbf{k}}}\right]$ for $\Phi_\mathbf{k}=\begin{pmatrix}
    M\Phi_{1\mathbf{k}} & \sqrt{Mm}\Phi_{2\mathbf{k}}\\
    \sqrt{Mm}\Phi_{2\mathbf{k}} & m\Phi_{3\mathbf{k}}
\end{pmatrix}$ and $\varepsilon^\pm_\mathbf{k}=\sqrt{\frac{\Phi_{1\mathbf{k}}+\Phi_{3\mathbf{k}}}{2}\pm\sqrt{\left(\frac{\Phi_{1\mathbf{k}}-\Phi_{3\mathbf{k}}}{2}\right)^2+\Phi^2_{2\mathbf{k}}}}$ the optical (acoustic) phonon energy dispersion.
In this phonon creation and annihilation basis, the magnon-phonon coupling is written as
\begin{align}
    H_{mp}&=\frac{2D_0\tan{\frac{\theta}{2}}\sqrt{2S^3}}{a}\sum_\mathbf{k}\sum_{\diamond{ilmj}}\left[\left(\hat{\mathbf{R}}^0_{ij}\cdot\hat{\mathbf{y}}\right)\left(a^\dagger_\mathbf{k}-a^{}_{-\mathbf{k}}\right)-i\left(\hat{\mathbf{R}}^0_{ij}\cdot\hat{\mathbf{x}}\right)\left(a^\dagger_\mathbf{k}+a^{}_{-\mathbf{k}}\right)\right]\nonumber\\
&\times\left[\left(\frac{\bar{b}^\dagger_{-\mathbf{k}}+\bar{b}^{}_\mathbf{k}}{\sqrt{2M\varepsilon^-_\mathbf{k}}}\cos\theta_\mathbf{k}-\frac{b^\dagger_{-\mathbf{k}}+b^{}_\mathbf{k}}{\sqrt{2M\varepsilon^+_\mathbf{k}}}\sin\theta_\mathbf{k}\right)\sin{\mathbf{k}_{ij}}-2\left(\frac{\bar{b}^\dagger_{-\mathbf{k}}+\bar{b}^{}_\mathbf{k}}{\sqrt{2m\varepsilon^-_\mathbf{k}}}\sin\theta_\mathbf{k}+\frac{b^\dagger_{-\mathbf{k}}+b^{}_\mathbf{k}}{\sqrt{2m\varepsilon^+_\mathbf{k}}}\cos\theta_\mathbf{k}\right)\sin{\frac{\mathbf{k}_{ij}}{2}}\cos{\frac{\mathbf{k}_{lm}}{2}}\right],
\end{align}
Then the full Hamiltonian can be expressed in the magnon and phonon creation and annihilation basis as
\begin{align}
    H=\sum_\mathbf{k}\left[\varepsilon^m_\mathbf{k}a^\dagger_\mathbf{k}a^{}_\mathbf{k}+\varepsilon^+_\mathbf{k}b^\dagger_\mathbf{k}b^{}_\mathbf{k}+\varepsilon^-_\mathbf{k}\bar{b}^\dagger_\mathbf{k}\bar{b}^{}_\mathbf{k}+\left({V^+_{\mathbf{k}}}^\dagger a^\dagger_\mathbf{k}b^{}_\mathbf{k}+{V^-_{\mathbf{k}}}^\dagger a^\dagger_\mathbf{k}\bar{b}^{}_\mathbf{k}+h.c.\right)+\hat{\mathcal{V}}_\mathbf{k}\right],
\end{align}
where $V^{+(-)}_\mathbf{k}$ is the coupling between magnons and optical (acoustic) phonons that conserves the quasiparticle number,
\begin{align}
V^{+}_\mathbf{k}&=-\frac{2D_0\tan{\frac{\theta}{2}}\sqrt{2S^3}}{a}\sum_{\diamond{ilmj}}i\left[\hat{\mathbf{R}}^0_{ij}\cdot\left(\hat{\mathbf{x}}-i\hat{\mathbf{y}}\right)\right]\left(\frac{2\cos\theta_\mathbf{k}\sin{\frac{\mathbf{k}_{ij}}{2}}\cos{\frac{\mathbf{k}_{lm}}{2}}}{\sqrt{2m\varepsilon^+_\mathbf{k}}}+\frac{\sin\theta_\mathbf{k}\sin{\mathbf{k}_{ij}}}{\sqrt{2M\varepsilon^+_\mathbf{k}}}\right),\\
V^{-}_\mathbf{k}&=-\frac{2D_0\tan{\frac{\theta}{2}}\sqrt{2S^3}}{a}\sum_{\diamond{ilmj}}i\left[\hat{\mathbf{R}}^0_{ij}\cdot\left(\hat{\mathbf{x}}-i\hat{\mathbf{y}}\right)\right]\left(\frac{2\sin\theta_\mathbf{k}\sin{\frac{\mathbf{k}_{ij}}{2}}\cos{\frac{\mathbf{k}_{lm}}{2}}}{\sqrt{2m\varepsilon^-_\mathbf{k}}}-\frac{\cos\theta_\mathbf{k}\sin{\mathbf{k}_{ij}}}{\sqrt{2M\varepsilon^-_\mathbf{k}}}\right),
\end{align}
and $\hat{\mathcal{V}}_\mathbf{k}$ includes all other quasiparticle-number non-conserving terms. As the physics studied in this work mainly results from the anti-crossings between magnons and phonons, and these gaps are opened predominantly by the quasiparticle-number conserving terms, we can thus safely ignore $\hat{\mathcal{V}}_\mathbf{k}$. Furthermore, since in most cases the acoustic branch and optical branch are well-separated with each other, the effects of $V^{-(+)}_\mathbf{k}$ is perturbatively small when we focus on the anti-crossings between magnons and optical (acoustic) phonons, and thus we end up with an effective two-band model Hamiltonian as
\begin{align}
    \tilde{H}^{\pm}_\mathbf{k}=\begin{pmatrix}
        \varepsilon^m_\mathbf{k} & {V^{\pm}_\mathbf{k}}^\dagger\\
        V^{\pm}_\mathbf{k} & \varepsilon^\pm_\mathbf{k}
    \end{pmatrix}=\frac{1}{2}\left(\varepsilon^m_\mathbf{k}+\varepsilon^\pm_\mathbf{k}\right)I_2+\mathbf{d}_\mathbf{k}^\pm\cdot\boldsymbol{\sigma},\text{ with }\mathbf{d}_\mathbf{k}^\pm=\left(\text{Re}[V^\pm_\mathbf{k}],\text{Im}[V^\pm_\mathbf{k}],\frac{\varepsilon^m_\mathbf{k}-\varepsilon^\pm_\mathbf{k}}{2}\right).
\end{align}
More explicitly, for the square lattice example, $\hat{\mathbf{R}}^0_{ij}=\{\hat{\mathbf{x}},\hat{\mathbf{y}}\}$, and the dynamical matrix of Eq.~(6) can be easily derived as,
\begin{align}
    \Phi(\mathbf{k})&=4\tilde{M}\omega^2_{uv}\begin{pmatrix}
        1 & -\cos\frac{k_x a}{2}\cos{\frac{k_ya}{2}}\\
        -\cos\frac{k_x a}{2}\cos{\frac{k_ya}{2}} & 1
    \end{pmatrix}\nonumber\\
    &+2M\omega^2_{u}\begin{pmatrix}
        2-\cos{(k_xa)}-\cos{(k_ya)} & 0\\
        0 & 0
    \end{pmatrix},
\end{align}
with $\tilde{M}\equiv\frac{2mM}{m+M}$ as the reduced mass, $\omega_{uv}=\sqrt{k_{uv}/\tilde{M}}$ and $\omega_u=\sqrt{k_u/M}$ two intrinsic frequencies. Substituting Eq.~(S17), Eq.~(S18) and Eq.~(S20) into Eq.~(S19), we have the effective model Eq.~(11) in the main text.
\section{Uncoupled magnon and phonon bands}
\begin{figure}[ht]
\centering
\includegraphics[width=0.95\textwidth]{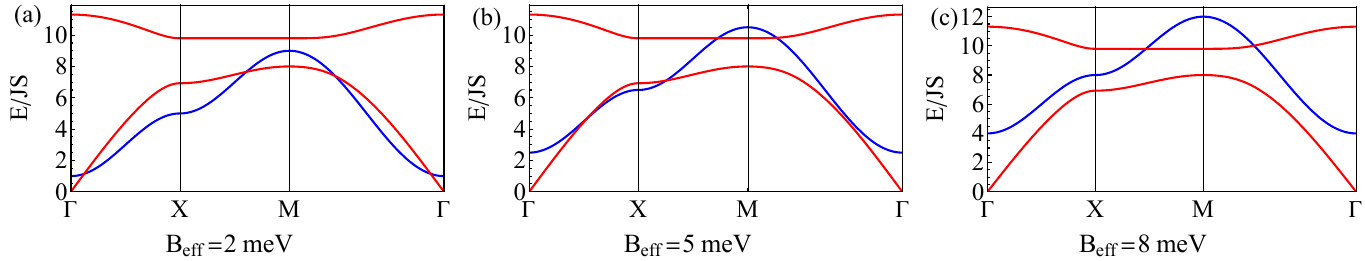}
 \caption{Uncoupled magnon and phonon band structure. The red (blue) lines show the phonon (magnon) dispersion for (a) $B_\text{eff}=2$ meV, (b) $B_\text{eff}=5$ meV. (c) $B_\text{eff}=8$ meV. Other parameters are the same as those in Fig.~2.} 
    \label{fig:uncoupled}
\end{figure}

In Fig.~\ref{fig:uncoupled}, we provide uncoupled magnon and phonon dispersion with $D_0=0$.
\section{Dynamical multiferroicity and toroidal moments}
The toroidal moment $\mathbf{T}$ is defined as the cross product between electrical polarization $\mathbf{P}$ and magnetic moment $\mathbf{M}$, i.e., $\mathbf{T}=\mathbf{P}\times\mathbf{M}$. Statically, $\mathbf{T}=0$ as $\mathbf{P}=0$, while dynamically $\mathbf{T}=\Delta\mathbf{P}\times\mathbf{M}$ can be finite. More explicitly, 
\begin{align}
    \mathbf{T}&=\sum_{\langle ij \rangle}\Delta\mathbf{P}_{ij}\times\mathbf{M}_i=\sum_{i}\left[Q^*(u^z_i-\frac{1}{4}\sum_{j\in n.n.}v^z_j)\hat{\mathbf{z}}\times\gamma\hbar\mathbf{S}_i\right]\equiv\sum_\mathbf{k}\mathbf{T}_\mathbf{k}\\
    &\approx\frac{1}{4}\hbar\gamma Q^*\sqrt{\frac{S}{2}}\sum_\mathbf{k}\sum_{\mathbf{r}_j\in n.n.}\left(u^z_\mathbf{k}-v^z_\mathbf{k}e^{i\mathbf{k}\cdot\mathbf{r}_j}\right)\left[i\left(a^{}_{-\mathbf{k}}-a^\dagger_\mathbf{k}\right)\hat{\mathbf{x}}+\left(a^{}_{-\mathbf{k}}+a^\dagger_{\mathbf{k}}\right)\hat{\mathbf{y}}\right]\\
    &=\frac{1}{4}\hbar\gamma Q^*\sqrt{\frac{S}{2}}\sum_\mathbf{k}\sum_{\mathbf{r}_j\in n.n.}\left[u^z_\mathbf{k}-v^z_\mathbf{k}\cos{\left(\mathbf{k}\cdot\mathbf{r}_j\right)}\right]\left[i\left(a^{}_{-\mathbf{k}}-a^\dagger_\mathbf{k}\right)\hat{\mathbf{x}}+\left(a^{}_{-\mathbf{k}}+a^\dagger_{\mathbf{k}}\right)\hat{\mathbf{y}}\right],
\end{align}
with $\pm Q^*$ the (total) effective charge of the cation (surrounding anions) at site $i$, $\gamma$ the gyromagnetic ratio, and $\mathbf{T}_\mathbf{k}$ the momentum-resolved toroidal moment. The last equation is because $\mathbf{P}=0$ and the surrounding anions are inversional symmetric with respect to site $i$. For the square lattice example in the main-text,
\begin{align}
    \mathbf{T}_\mathbf{k}\approx\hbar\gamma Q^*\sqrt{\frac{S}{2}}\left(u^z_\mathbf{k}-v^z_\mathbf{k}\cos{\frac{k_x a}{2}}\cos{\frac{k_y a}{2}}\right)\left[i\left(a^{}_{-\mathbf{k}}-a^\dagger_\mathbf{k}\right)\hat{\mathbf{x}}+\left(a^{}_{-\mathbf{k}}+a^\dagger_{\mathbf{k}}\right)\hat{\mathbf{y}}\right]
\end{align}

As the effects of dynamical DMI are more significant when magnons couple with optical phonons, we focus on the two-band subspace that consists of the basis $
\tilde{\mathbf{X}}_\mathbf{k}=(a_\mathbf{k},b_\mathbf{k})^T$. It is well studied that the two eigenstates of such a two-band Hamiltonian $\tilde{H}^+_\mathbf{k}$ can be expressed as
\begin{align}
    \left|\psi^\pm_\mathbf{k}\right\rangle=\frac{1}{\sqrt{2|\mathbf{d}_\mathbf{k}^+|\left(|\mathbf{d}_\mathbf{k}^+|\pm d_{z,\mathbf{k}}^+\right)}}\begin{pmatrix}
        d_{z,\mathbf{k}}^+\pm |\mathbf{d}_{\mathbf{k}}^+|\\
        d_{x,\mathbf{k}}^++id_{y,\mathbf{k}}^+
    \end{pmatrix}
\end{align}
in the basis of $\tilde{\mathbf{X}}_\mathbf{k}$.

The $\mathbf{k}$-resolved toroidal moment for these two states reads as
\begin{align}
    \langle\mathbf{T_\mathbf{k}}\rangle_\pm=\langle \psi^\pm_\mathbf{k}|\mathbf{T}_\mathbf{k}|\psi^\pm_\mathbf{k}\rangle=\mp\frac{1}{4}\left[\hbar\gamma Q^*\sqrt{\frac{S}{2}}\sum_{\mathbf{r}_j\in n.n.}\left(\frac{\sin\theta_\mathbf{k}}{\sqrt{2M\varepsilon^+_\mathbf{k}}}+\frac{\cos\theta_\mathbf{k}\cos{\left(\mathbf{k}\cdot\mathbf{r}_j\right)}}{\sqrt{2m\varepsilon^+_\mathbf{k}}}\right)\right]\left(\hat{d}^+_{y,\mathbf{k}}\hat{\mathbf{x}}+\hat{d}^+_{x,\mathbf{k}}\hat{\mathbf{y}}\right),
\end{align}
where $\hat{d}^+_{\alpha,\mathbf{k}}=d^+_{\alpha,\mathbf{k}}/|\mathbf{d}^+_\mathbf{k}|$ for $\alpha=x,y,z$.

Since the Berry curvature for $|\psi^\pm_\mathbf{k}\rangle$ is $\Omega^\pm_\mathbf{k}=\pm\frac{1}{2}\hat{\mathbf{d}}^+_\mathbf{k}\cdot\left(\partial_{k_x}\hat{\mathbf{d}}^+_\mathbf{k}\times\partial_{k_y}\hat{\mathbf{d}}^+_\mathbf{k}\right)$, and $\hat{d}^+_{x,\mathbf{k}},\hat{d}^+_{y,\mathbf{k}}\propto D_0$, we find that the dynamical DMI, the induced band topology and dynamical toroidal (magnetoelectric) moments are all connected with each others.

\end{document}